# Measuring Strategy-Decay Risk: Minimum Regime Performance and the Durability of Systematic Investing

**Nolan Alexander and Frank J. Fabozzi**

**Nolan Alexander** is a PhD student in the Department of Systems Engineering in the School of Engineering and Applied Sciences at the University of Virginia. Charlottesville, VA
nka5we@virginia.edu

**Frank J. Fabozzi** is professor of practice in finance at Johns Hopkins University's Carey Business School. Baltimore, MD
ffabozz1@jhu.edu

## Key Takeaways

- Minimum Regime Performance (MRP) captures the weakest realized efficiency of a systematic strategy across market regimes, translating the abstract notion of robustness into a practical, observable risk metric.
- Empirical results reveal a "decay-risk frontier," showing the tradeoff between strategy efficiency and durability.
- Incorporating MRP into portfolio construction and oversight adds a temporal dimension to the risk taxonomy, enabling investors to manage not only how much risk they take, but how long their strategies remain effective.

## Abstract

Systematic investment strategies are exposed to a subtle but pervasive vulnerability: the progressive erosion of their effectiveness as market regimes change. Traditional risk measures, designed to capture volatility or drawdowns, overlook this form of structural fragility. This paper introduces a quantitative framework for assessing the durability of systematic strategies through Minimum Regime Performance (MRP), defined as the lowest realized risk-adjusted return across distinct historical regimes. MRP serves as a lower bound on a strategy's robustness, capturing how performance deteriorates when underlying relationships weaken or competitive pressures compress alpha. Applied to a broad universe of established factor strategies, the measure reveals a consistent trade-off between efficiency and resilience, strategies with higher long-term Sharpe ratios do not always exhibit higher MRPs. By translating the persistence of investment efficacy into a measurable quantity, the framework provides investors with a practical diagnostic for identifying and managing strategy-decay risk, a novel dimension of portfolio fragility that complements traditional measures of market and liquidity risk.

Systematic investing has become the dominant language of modern portfolio construction. Factor models, machine-learning systems, and rules-based strategies now guide capital allocation across virtually every asset class. Yet as these models proliferate, they introduce a new and often underappreciated form of vulnerability: the gradual erosion of performance as the underlying relationships that support a strategy weaken, become crowded, or disappear. This phenomenon—commonly referred to as *alpha decay*—is not a transitory fluctuation in returns but a structural breakdown of the investment process itself. Traditional risk measures such as volatility, value at risk, or drawdown do not capture this process because they describe the variability of outcomes, not the degradation of *efficacy*.

The persistence problem has been widely documented. McLean and Pontiff (2016) show that academic factors experience substantial performance deterioration after publication, suggesting that awareness and adoption erode returns. Harvey, Liu, and Zhu (2016) argue that the proliferation of published factors reflects data mining as much as discovery, compounding the risk of overfitted models. More broadly, Bailey and López de Prado (2014) demonstrate that backtest overfitting can lead investors to overestimate the robustness of systematic strategies even before they are deployed. Together, these findings point to a pervasive meta-risk: the risk that a process built on empirical regularities loses validity as markets adapt.

This paper introduces Minimum Regime Performance (MRP) as a diagnostic for quantifying that meta-risk, here termed strategy-decay risk. The concept is straightforward: if a strategy's risk-adjusted performance is evaluated separately across distinct market regimes, the *minimum* of those regime-level performances represents a natural measure of robustness. MRP captures the worst realized Sharpe ratio across structural environments, interpreting that minimum as a lower bound on the durability of the investment process. Whereas volatility and drawdown measure how much a portfolio can lose, MRP measures *how far a strategy's efficacy can fall* when its supporting relationships weaken. The MRP framework aligns conceptually with the emerging literature on model fragility and process risk. Cont (2016) and Farmer et al. (2021) argue that nonstationarity in market dynamics demands a shift from static to adaptive risk assessment. Kritzman and Li (2010) introduce the concept of model-based instability as a distinct source of portfolio risk, separate from return volatility. Simonian (2022) extends this notion by characterizing model risk as the uncertainty in an investment process's ability to remain valid under changing conditions. The proposed metric operationalizes these ideas by providing a single interpretable statistic that can be computed from realized data and monitored over time.

We apply the framework to a universe of well-established factor strategies, following the dataset constructed by Jensen, Kelly, and Pedersen (2023). These factors provide an ideal test bed because their economic logic is clear, their crowding dynamics are well understood, and their post-publication decay has been documented in numerous studies. By comparing the full-sample Sharpe ratios of these factors with their MRPs, we illustrate how traditional performance evaluation can mask fragility. The empirical implementation, along with sensitivity tests for regime definitions and sampling frequency, is presented in Appendix B.

The remainder of the paper proceeds as follows. The next section outlines the conceptual framework behind MRP, showing how it reframes performance evaluation in terms of robustness rather than central tendency. The subsequent section presents empirical illustrations using cross-sectional factors to demonstrate the existence of a decay-risk frontier—the trade-off between efficiency and durability. The final section discusses portfolio implications and situates strategy-decay risk within the broader taxonomy of *novel risks*, alongside liquidity, counterparty, and model risk. Together, these elements provide a foundation for extending risk management beyond outcomes to the longevity of the investment process itself.

## CONCEPTUAL FRAMEWORK

The concept of strategy-decay risk begins with a simple observation: performance statistics derived from long samples implicitly assume that a strategy's underlying relationships are stable. In practice, the return-generating process of most systematic strategies is not stationary. Correlations shift, market microstructure evolves, and behavioral inefficiencies become arbitraged away. As a result, the performance observed in one environment may not generalize to the next. The relevant question for allocators is therefore not, *What is the average Sharpe ratio?* but rather, *How low can the Sharpe ratio fall when the environment changes?*

To answer this, the $\text{MRP}_s$ framework divides the return history of a strategy into a set of $s + 1$ distinct market regimes, each representing a structural environment with internally consistent dynamics. Most applications of MRP will only require one split, i.e. $s = 1$ to model simpler two-regime structures. Let $x$ be a return series that will be split into $s$ subsets or regimes $\{r_1, r_2, \dots, r_s\}$. Within each regime, a standard risk-adjusted performance measure such as the Sharpe ratio $S$ or information ratio $I$ is calculated. The metric then searches all possible regime split choices such that each regime is at least length $d$, and selects the regime splits that yield the minimum performance measure. The MRP is then defined as the minimum value across all selected regimes.

To motivate the general formulation, we start by formally defining the MRP with one split, the simplest instance of the framework. Let $x$ with a subscript denote a subset the return series corresponding to a regime. In the subscript, the value before the colon is the starting index, and the value after the colon is the ending index. With one split, $r_1 = x_{0:t_1-1}$ and $r_2 = x_{t_1:T}$. To define MRP, we must first define the minimum metric across regimes at split $t$,

$$m_1(x, \{t\}) = \min\big(S(r_1), S(r_2)\big). \tag{1}$$

The $\text{MRP}_1$ is the minimum $m_1$ across all possible splits:

$$\text{MRP}_1(x) = \min_{t \in [d, T-d]} m_1. \tag{2}$$

Now, we can define the generalized MRP for more than one split. Let $\mathcal{T} = \{t_1, t_2, \dots, t_s\}$ denote the set of $s$ splits. With $s$ splits, $r_1 = x_{0:t_1-1}$, $r_2 = x_{t_1:t_2-1}$, and $r_{s+1} = x_{t_s:T}$. The generalized version of Eq. 1, the minimum metric across regimes for splits $\mathcal{T}$, is

$$m_s(x, \mathcal{T}) = \min(S(r_1), S(r_2), \ldots, S(r_{s+1})). \quad (3)$$

The generalized version of Eq. 2, the MRP with $s$ splits, is

$$\text{MRP}_s(r) = \min_{\mathcal{T}} m_s. \quad (4)$$

This definition treats each regime as an independent realization of the strategy's efficacy. The lowest of these values represents the weakest environment in which the strategy has been observed to operate successfully. The analytical relationship between MRP, regime sampling frequency, and bias relative to the full-sample Sharpe ratio is demonstrated in Appendix A.

The intuition is analogous to expected shortfall in portfolio risk management. Whereas expected shortfall estimates the mean of the worst $q\%$ of returns, MRP identifies the worst *structural* performance—the lower tail of the strategy's risk-adjusted efficiency distribution. In that sense, MRP can be interpreted as the "expected shortfall of model efficacy." This perspective connects naturally to the literature on process fragility, which defines robustness not by the magnitude of average outcomes but by sensitivity to unfavorable conditions (Kritzman and Li 2010; Cont 2016).

MRP is in the same units as the the risk-adjusted performance metric calculated on a single regime, which will usually be the Sharpe ratio. This allows the metric to provide greater interpretability as portfolio managers often have intuition of the relative value of Sharpe ratios.

MRP is intentionally conservative. Because it selects the lowest realized regime performance, it will always underestimate the unconditional Sharpe ratio. This downward bias is a design feature rather than a flaw: it reflects a *robustness preference* similar to that embedded in stress testing or worst-case optimization (Ben-Tal and Nemirovski 1998). For investors allocating across many systematic strategies, this bias is desirable. It enables comparison of strategies on the basis of resilience rather than historical efficiency, mitigating the tendency to overweight high-Sharpe but fragile signals.

| Metric | Captures | Limitations | When It Works Best | Where MRP Adds Value |
|---|---|---|---|---|
| **Sharpe Ratio** | Average excess return per unit of total volatility. Widely understood and comparable across assets. | Assumes stationarity; averages over sample; treats upside and downside volatility equally. | Stable environments with constant risk/return. | MRP reveals whether Sharpe performance is concentrated in a few favorable regimes. |
| **Sortino Ratio** | Excess return relative to downside volatility only, penalizing harmful variation. | Still averages across sample; ignores regime shifts; sensitive to downside threshold choice. | Strategies with asymmetric return distributions. | MRP identifies persistence across subsamples rather than reliance on downside adjustment. |

| Metric | Captures | Limitations | When It Works Best | Where MRP Adds Value |
|---|---|---|---|---|
| **MRP** | Risk-adjusted performance across optimal subsamples; adapts to structural breaks and volatility shifts. | Sensitive to minimum window choice; requires more computation. | Non-stationary, crisis-prone, or regime-shifting environments. | Provides resilience insights: shows if performance is persistent or concentrated in tail periods. |

Exhibit 1. Comparison of Sharpe, Sortino, and MRP

The framework also extends to multi-strategy portfolios. Consider a vector of strategy returns $X_t = (x_{1t}, \ldots, x_{nt})$and a set of weights $w$. The portfolio's MRP can be expressed as the minimum Sharpe ratio across all historical regimes, computed using the weighted aggregate return $w'X_t$. The portfolio-level MRP is therefore sensitive to both strategy-level fragility and regime covariance. This property enables investors to assess the collective robustness of their allocations, identifying concentrations of decay risk that are analogous to factor concentration in conventional risk models.

Exhibit 2 illustrates the distinction between average performance and MRP using the Profit Growth factor. The figure shows the MRP calculated with one split applied to the historical return series. Although the full-sample Sharpe ratio appears stable, the minimum regime performance reveals a significant decline in efficiency after the split, corresponding to a structural shift in market conditions. The example illustrates how traditional average statistics can conceal a crucial aspect of risk, the potential for a profitable strategy to fail when the environment shifts.

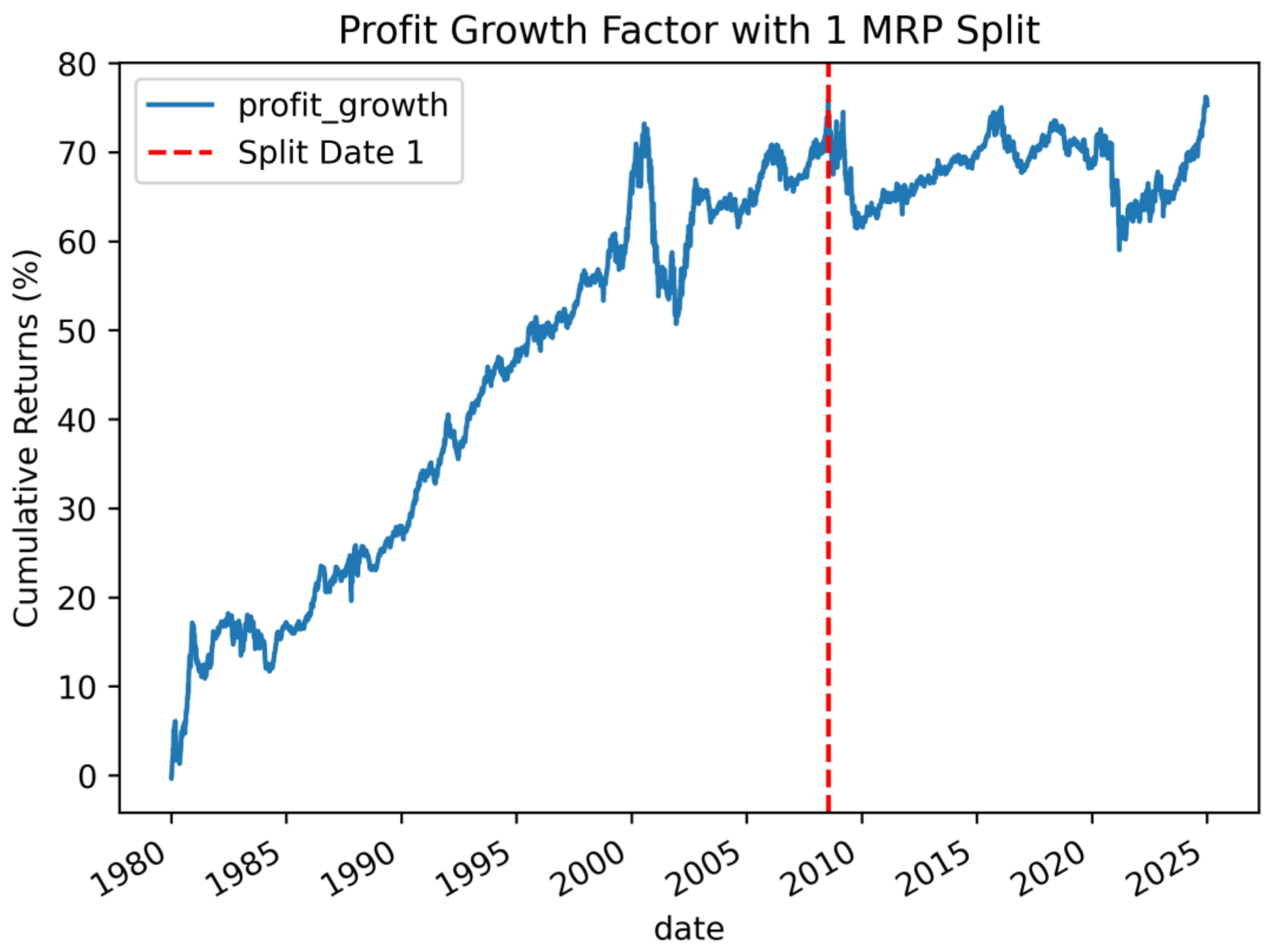

Exhibit 2. The MRP with One Split Applied to the Profit Growth Factor

MRP thereby reframes performance evaluation in terms of *temporal resilience*. Rather than asking whether a strategy is profitable on average, it asks whether the underlying process remains valid across environments. The measure is diagnostic rather than predictive: it does not forecast future decay but identifies the degree of historical vulnerability to structural breaks. Because it relies solely on realized data, it can be updated continuously as new regimes emerge, functioning as a real-time indicator of process stability.

From a governance standpoint, MRP offers two practical advantages. First, it provides a standardized statistic for comparing the *durability* of strategies across managers or asset classes, much as the Sharpe ratio standardizes efficiency. Second, it enables early detection of potential model breakdowns. A declining MRP over successive updates indicates that the strategy's weakest regime is worsening, a warning signal that warrants further review. These applications make MRP particularly relevant in the context of *model risk management*, where supervisors and allocators require interpretable metrics to monitor the health of the process over time.[1]

[1] See Simonian (2022) and Farmer et al. (2021).

The analytical properties of MRP—its sampling behavior, bias decomposition, and asymptotic limits under stationary and nonstationary processes—are formally demonstrated in Appendix A. The empirical illustrations in the next section show how the measure behaves when applied to real strategies and how it reveals the trade-off between efficiency and durability that defines strategy-decay risk.

## EMPIRICAL ILLUSTRATION

The practical relevance of MRP lies in its ability to identify *when* and *how* a systematic strategy becomes fragile. To demonstrate its behavior empirically, we apply the framework to a broad universe of published factor strategies whose long histories and well-documented performance patterns make them ideal laboratories for studying decay risk.

The analysis draws on the cross-sectional factors compiled by Jensen, Kelly, and Pedersen (2023), which encompass value, momentum, profitability, investment, quality, and low-risk styles across U.S. equities. Monthly excess returns are examined over the period 1965–2023. In our analysis, we start the factor data in 1980 to ensure that all factors are already live at the start of our analysis. Consistent with McLean and Pontiff (2016), these factors display a common life cycle: strong pre-publication performance followed by attenuation once the signal becomes widely adopted. Such a pattern provides a natural context in which to evaluate MRP as a measure of durability. Full data sources, preprocessing steps, and robustness checks for alternative definitions of regime boundaries are described in Appendix B.

To estimate MRP, each factor's return history is segmented into distinct regimes based on structural changes in market volatility, liquidity, and monetary policy. This segmentation follows the regime-identification framework in Farmer et al. (2021) and is designed to capture long-term shifts in macroeconomic and behavioral dynamics rather than short-term fluctuations. Within each regime, the annualized Sharpe ratio is computed, and the minimum value is recorded as the MRP.

Exhibit 3 summarizes the full-sample Sharpe ratios and MRPs for all factors using a 40 year lookback and *d* equal to two years. Two patterns emerge. First, the dispersion between average and minimum regime performance is substantial, with differences exceeding 0.50 Sharpe points for several factors. Second, some factors with the high relative historical Sharpe ratios—most notably Debt Issuance and Investment—exhibit the largest relative decline when evaluated through MRP. These results suggest that high apparent efficiency can coincide with structural fragility. However, this is not always the case. Quality has a high Sharpe, but does not decrease as significantly as other factors, and is the only factor to have a positive MRP value.

| **Factor Name** | **SR** | **$MRP_1$** | **Left SR** | **Right SR** | **Split Date** |
|---|---|---|---|---|---|
| Accruals | 0.74 | -0.09 | 0.81 | -0.09 | 09/23/21 |
| Debt Issuance | 1.14 | -0.05 | 1.38 | -0.05 | 12/07/16 |
| Investment | 0.45 | -1.01 | 0.49 | -1.01 | 12/29/22 |
| Low Leverage | 0.01 | -0.89 | -0.89 | 0.05 | 06/21/85 |
| Low Risk | 0.13 | -0.61 | 0.16 | -0.61 | 12/29/22 |
| Momentum | 0.37 | -0.08 | 0.39 | -0.08 | 11/10/22 |
| Profit Growth | 0.46 | -0.01 | 0.76 | -0.01 | 07/16/08 |
| Profitability | 0.49 | -0.25 | -0.25 | 0.54 | 03/28/84 |
| Quality | 0.65 | 0.06 | 0.74 | 0.06 | 11/09/20 |
| Seasonality | 0.70 | -0.22 | 0.91 | -0.22 | 06/28/16 |
| Short Term Rev | 0.31 | -0.11 | 0.35 | -0.11 | 07/16/21 |
| Size | 0.16 | -0.93 | 0.3 | -0.93 | 03/15/21 |
| Value | 0.37 | -0.41 | 0.4 | -0.41 | 12/28/22 |

Exhibit 3. Summary of Full-Sample Sharpe Ratios and Minimum Regime Performances (MRPs) for All Factors.

The relationship between average Sharpe ratio and MRP is visualized in Exhibit 4. Each point represents a factor strategy, plotted by its full-sample Sharpe ratio on the horizontal axis and its MRP on the vertical axis. The resulting “decay-risk frontier” resembles the mean–variance frontier familiar to portfolio managers: investors must choose between higher expected efficiency and greater durability. The decay-risk frontier show that multiple factors are dominated by other factors with both higher MRP and Sharpe. Some of the most dominated factors include Low Leverage, Size, Low Risk, and Investment.

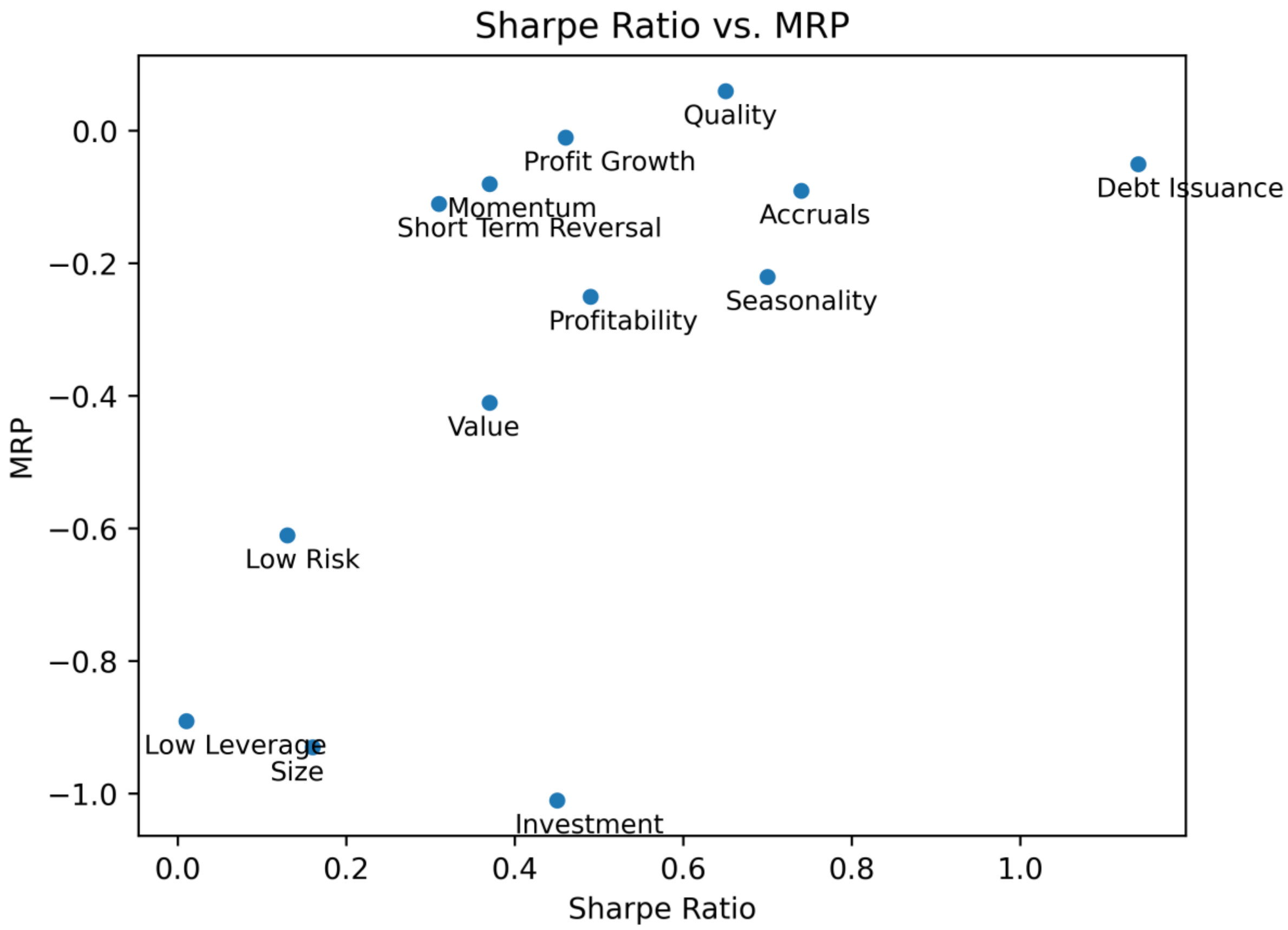

Exhibit 4. Sharpe Ratio versus Minimum Regime Performance across All Factors

**Sensitivity Analysis of MRP**

To assess whether the results are sensitive to methodological choices, we perform a robustness analysis of the MRP estimates with respect to two parameters: the look-back length used for regime segmentation and the minimum-sample parameter $d$, which determines the number of observations required to compute a regime-specific Sharpe ratio.

The full results are summarized in Exhibits 5 and 6. Detailed numerical values for each factor appear in the supplementary material available upon request.

Exhibit 5 reports the sensitivity of MRP minus Sharpe across alternative look-back lengths. This heatmap shows the average MRP minus Sharpe ratio value of 1 to 5 years of values of $d$ separated by lookback lengths of 10 to 40 years. The behavior of the lookback on MRP value is dependent on the specific time-series as going further back in time can reveal new regimes. Generally, we see that as the lookback increases, the MRP minus Sharpe decreases. The mean and median MRPs remain largely stable across parameterizations, confirming that the framework's identification of fragile versus resilient factors is not an artifact of the sampling frequency.

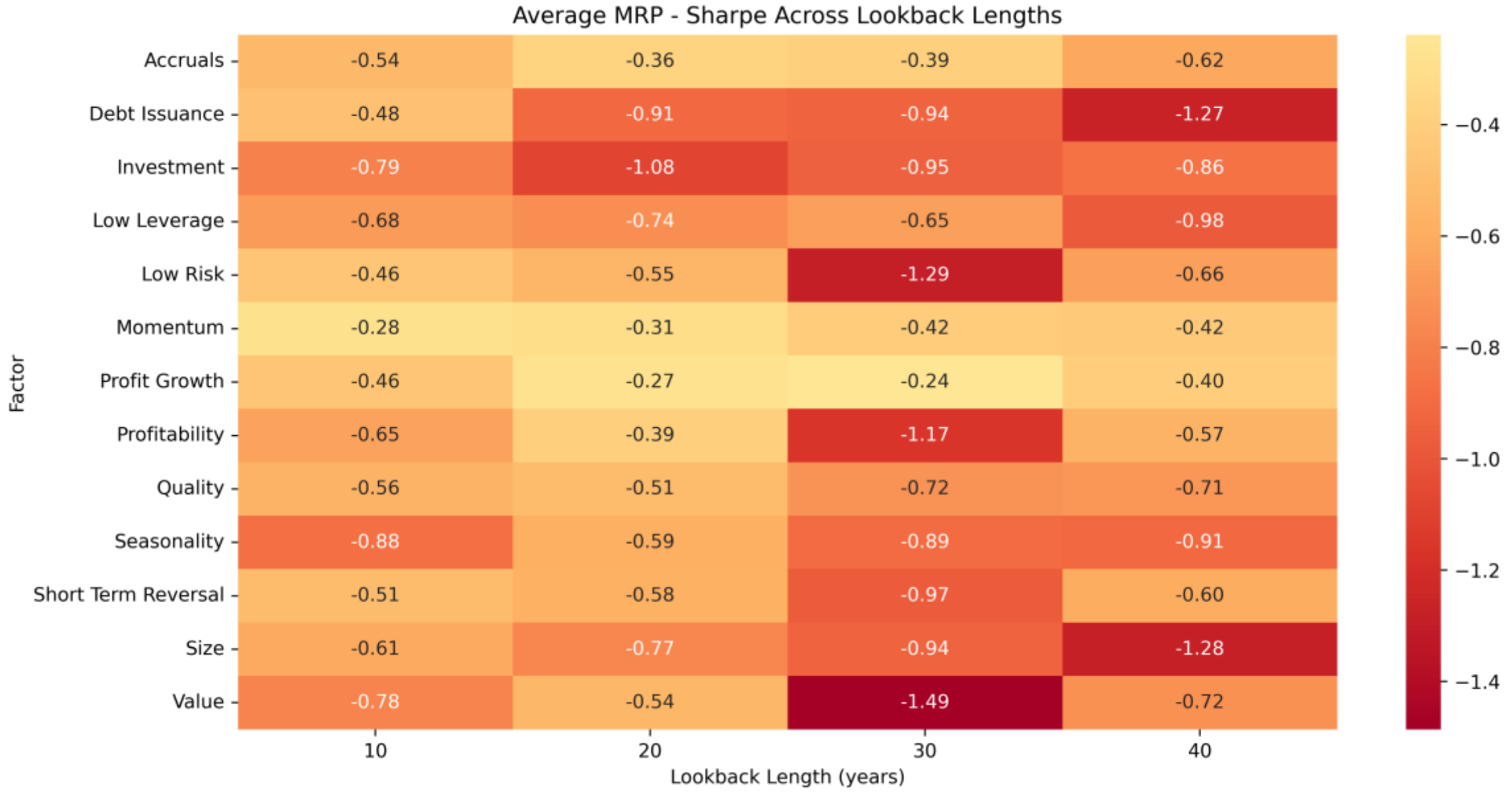

Exhibit 5. MRP – Sharpe Sensitivity by Look-Back Length

Exhibit 6 shows the sensitivity of MRP minus Sharpe values to the minimum-sample parameter *d*. This heatmap shows the average MRP minus Sharpe values of lookback lengths of 10 to 40 years separated by values of *d* from 1 to 5 years. As *d* increases, regimes with fewer observations are excluded from the computation, reducing the influence of short or noisy intervals. However, this requires an assumption that regimes must persist for a longer period of time, which on average increases the MRP value. Regardless, the cross-sectional ranking of factors remains highly consistent across all look-back settings, indicating that the decay-risk ordering observed earlier is robust to sample-size thresholds.

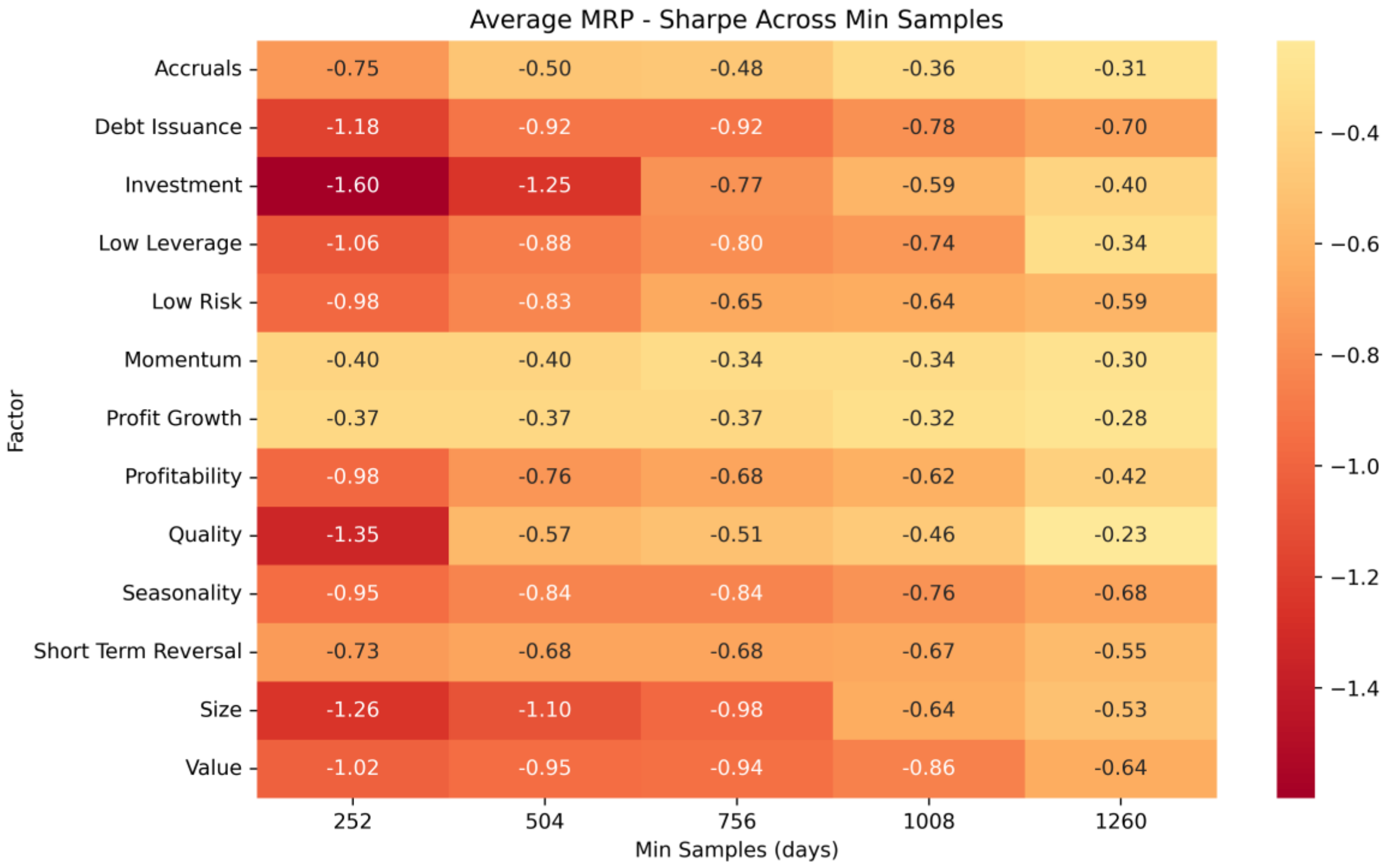

Exhibit 6: MRP – Sharpe Sensitivity by Minimum-Sample Parameter *d*

Together, these results demonstrate that the MRP framework captures a persistent and economically meaningful dimension of fragility. Across specifications, the strategies identified earlier as structurally vulnerable, such as Investment and Size, retain low MRPs, while defensive styles such as Quality and Accruals remain durable.

This stability supports the interpretation of MRP as a genuine measure of strategy-decay risk rather than a statistical artifact of regime definition or sample selection.

To ensure that MRP captures a distinct dimension of risk, we examine its relationship with other common robustness metrics. Exhibit 7 reports the cross-sectional correlations between MRP, full-sample Sharpe ratio, rolling-window Sharpe volatility, and maximum drawdown. The results show that MRP is only weakly correlated with volatility- and drawdown-based measures, confirming that it isolates an orthogonal aspect of fragility: the temporal durability of a strategy's efficacy rather than the variability of its returns.

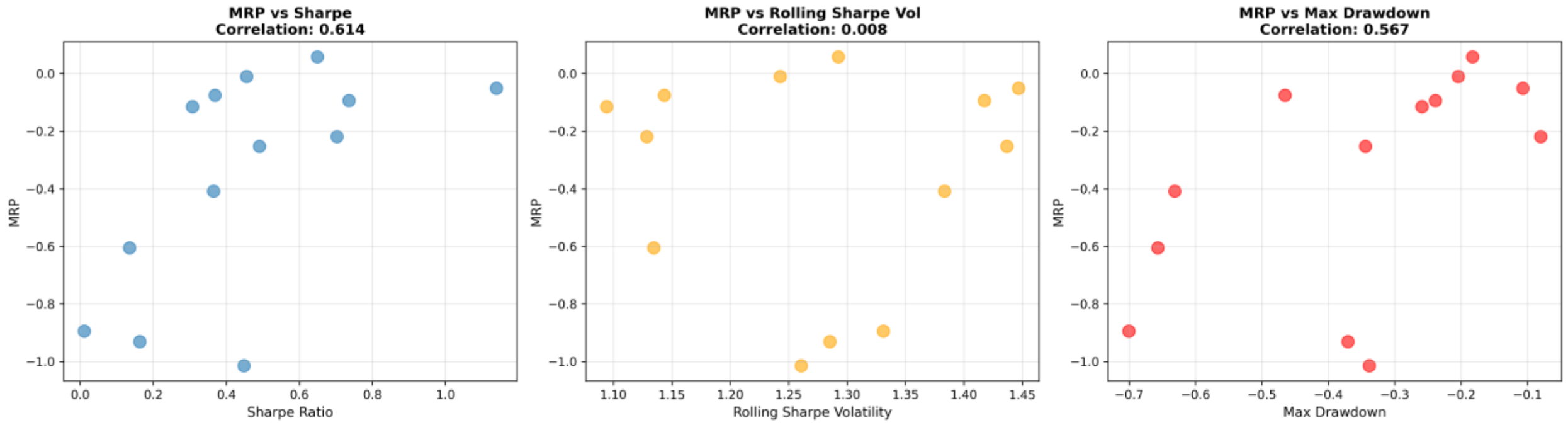

Exhibit 7. Cross-Sectional Correlations between Minimum Regime Performance (MRP) and Other Robustness Metrics.

These findings have several implications. First, regime sensitivity is pervasive—even strategies grounded in strong economic intuition exhibit periods of structural weakness. Second, the degree of decay risk varies significantly across different factors, with crowd-sensitive signals, such as momentum, exhibiting the most pronounced vulnerability. Third, MRP provides an interpretable and stable diagnostic that can be updated continuously as new data accrues, offering a real-time indicator of model fatigue.

From an allocator's perspective, the decay-risk frontier provides a clear and actionable visualization of resilience. Managers and strategies can be ranked not solely by average efficiency but by the ratio of MRP to full-sample Sharpe ratio, a simple index of robustness. Portfolios that overweight strategies with higher MRPs tend to experience smaller drawdowns during regime transitions and greater persistence in realized performance. Appendix B provides sensitivity tests using alternative regime definitions, subperiod analyses, and Monte Carlo resampling to confirm that these results are not artifacts of sample segmentation.

## INTERPRETATION AND PORTFOLIO IMPLICATIONS

The empirical results presented earlier reveal that systematic strategies vary widely in their resilience to changing regimes. Some, such as Quality and Accruals, exhibit high durability, while others, like Investment and Size, show steep performance erosion when market conditions shift. These findings underscore a broader insight: even when strategies are properly implemented and well diversified across factors, portfolios remain exposed to strategy-decay risk (i.e., the possibility that the alpha-generating mechanisms themselves lose relevance over time).

This form of risk differs fundamentally from traditional measures of portfolio volatility or drawdown. Volatility quantifies the dispersion of realized outcomes, while strategy-decay risk captures the deterioration in the validity of the process that produces those outcomes. In effect, MRP transforms robustness, often treated as a qualitative judgment, into a measurable quantity. By interpreting the lowest observed Sharpe ratio across regimes as a lower bound on durability, MRP converts the elusive notion of persistence into an observable statistic.

From a portfolio-construction standpoint, this opens a new axis of optimization: the trade-off between efficiency and resilience. The decay-risk frontier, as displayed in Exhibit 3, formalizes this relationship. Strategies that lie below the diagonal of parity are efficient but fragile; those closer to the frontier's upper boundary offer lower expected efficiency but greater robustness. Strategies that are in the top right corner are preferable to those towards the bottom and left. Allocators can use this information by limiting allocations to strategies whose minimum regime performance exceeds a threshold relative to their full-sample Sharpe ratio. MRP cannot be easily incorporated as a formal constraint to a portfolio optimizer because it is nonlinear, but MRP can be used in heuristic methods.

Beyond allocation, MRP serves as a governance and monitoring tool. Because it can be recalculated as new data become available, a declining MRP trajectory signals increasing fragility in a strategy's weakest regime. Such a trend can trigger review or recalibration before losses occur, analogous to how rising tracking error or declining information ratios prompt manager reassessment. In practice, investment committees can integrate MRP reporting into periodic risk reviews alongside volatility, liquidity, and factor-exposure reports. The interpretability of the measure makes it well-suited for communication between quantitative research teams and fiduciaries responsible for oversight.

The broader implication of these findings is conceptual. MRP adds a temporal dimension to the risk taxonomy that underpins systematic investing. Market, liquidity, and credit risks describe exposures to states of the world; model and process risks describe vulnerabilities in the tools used to interpret those states. Strategy-decay risk, as captured by MRP, bridges these categories—it measures how the interaction between models and markets evolves through time. This perspective aligns with recent thinking in adaptive market dynamics—such as the framework proposed by Lo, which highlights evolving, context-dependent efficiency—and underscores the growing consensus that monitoring the robustness of the investment process itself, in addition to its traditional outputs, is increasingly necessary (see also Farmer et al. 2021; Simonian 2022).

Finally, the operational burden of adopting MRP is minimal. The computation requires only regime segmentation, return data, and standard risk metrics. The conceptual shift, however, is significant. It encourages allocators to treat *process durability* as an asset-class–agnostic dimension of risk, one that should be monitored, budgeted, and diversified alongside volatility and correlation.

By transforming the persistence of alpha into measurable statistics, MRP allows practitioners to view robustness not as an afterthought but as an integral part of portfolio design. In doing so, it helps bridge the gap between traditional performance evaluation and the emerging discipline of resilience-based risk management that lies at the heart of this special issue on novel risks.

## CONCLUSION

Systematic investing has transformed the practice of portfolio management, but it has also created new sources of fragility. As models proliferate and signals become widely known, their performance often weakens, a process that traditional risk measures fail to detect. The framework introduced in this paper provides a direct method for measuring vulnerability. By focusing on the weakest realized performance across structural regimes, MRP provides an interpretable and implementable measure of strategy-decay risk.

Empirical evidence demonstrates that this risk is pervasive. Across a broad set of factor strategies, the dispersion between full-sample Sharpe ratios and their corresponding MRPs is substantial. Strategies that appear most efficient on average frequently prove least durable when market conditions shift. The resulting decay-risk frontier mirrors the trade-off between return and volatility but in the dimension of temporal stability. This finding suggests that investors must decide not only how much risk to take but also how much fragility to tolerate.

The practical implications are clear. MRP can be incorporated into portfolio construction as a robustness constraint or weighting adjustment, guiding allocations toward strategies that remain effective across environments. It can also serve as a monitoring metric within governance frameworks, allowing risk managers to detect early signs of model fatigue. By quantifying the durability of investment processes, MRP turns the abstract idea of "robustness" into an observable statistic that can be tracked over time.

Conceptually, MRP extends the modern risk taxonomy beyond the dispersion of returns to the persistence of process validity. In doing so, it addresses one of the defining challenges of the data-driven era: the risk that yesterday's model ceases to explain today's markets. Just as portfolio theory evolved to include liquidity, model, and counterparty risk, the inclusion of *strategy-decay risk* acknowledges that the stability of investment processes is itself a critical dimension of portfolio resilience. Measuring and managing that risk is essential not only for sustaining performance but for ensuring that systematic investing remains adaptive, transparent, and durable in an ever-changing market landscape.

# Appendix A: Statistical Properties of $\mathbf{MRP}_s$

We calculate the bias of MRP and show that it is an inconsistent estimator of the performance metric, which will usually be the Sharpe ratio. For the bias, we calculate both the exact integral form of the bias and an asymptotic approximation.

## Bias of $\mathbf{MRP}_s$

### Derivation of the Number of Valid Splits

To calculate the bias of $\text{MRP}_s$, we need to derive $n_s$, the number of valid splits of $r$. We have $|r| = n$, $s$ splits creating $s + 1$ segments of at least length $d$.

The minimum total length needed is $(s + 1) \cdot d$. Therefore, we require:

$$n \geq (s + 1)d$$

After allocating the minimum length $d$ to each of the $s + 1$ segments, we have $n - (s + 1)d$ units remaining to distribute freely among the segments. This is equivalent to a "stars and bars" problem: distributing $n - (s + 1)d$ identical items into $s + 1$ bins.

The number of ways to do this is given by the binomial coefficient:

$$n_s = \binom{n - (s + 1)d + (s + 1) - 1}{(s + 1) - 1}$$

Simplifying yields

$$n_s = \binom{n - sd - d + s}{s}, \tag{5}$$

which is the number of valid splits. This formula is valid only when $n \geq (s + 1)d$.

### Integral Expression of Bias

We will assume the Sharpe of the split segments are independent of one another to simplify the calculation of the bias. Let $X_i \sim \mathcal{N}(\mu, \sigma^2)$ be i.i.d. random variables representing $S(r)$. Let $n_s$ denote the number of valid splits of $r$, where each split is length at least $d$. Let $Y$ be a random variable representing $m_1$ defined as

$$Y_i = \min_{j \in [1,s]} (X_{ij}) \quad i \in [1, n_s]$$

.

Let $Z$ be a random variable representing $\mathrm{MRP}_1$ defined as

$$Z = \min(Y_1, Y_2, \dots, Y_{n_s}).$$

The bias of the $\mathrm{MRP}_1$ estimator is

$$b_{\mathrm{MRP}_1} = \mathbb{E}[X] - \mathbb{E}[Z] \tag{6}$$

This bias term is what we aim to calculate.

We will begin by standardizing. We define the standardized variables

$$X_{ij}{}' = \frac{X_{ij} - \mu}{\sigma} \sim \mathcal{N}(0,1),$$

such that

$$Y_i{}' = \min_{j\in[1,s]}(X_{ij}{}'), \quad Z' = \min(Y_1{}', \dots, Y_n{}').$$

By the linearity of scaling and translation for order statistics,

$$Z = \mu + \sigma Z',$$

and therefore,

$$\mathbb{E}[Z] = \mu + \sigma\, \mathbb{E}[Z']. \tag{7}$$

From the standard normal case,

$$Z' \overset{d}{=} \min(X_{1,1}{}', X_{1,2}{}', \dots, X_{2,1}{}', X_{2,2}{}', \dots, X_{n_s s}{}'),$$

That is $Z'$ has the same distribution as the minimum of $2n$ i.i.d. standard normal variables. Thus for the standard normal distribution, this is the integral form:

$$\mathbb{E}[Z'] = sn_s \int_{-\infty}^{\infty} z\ \phi(z)\,[1 - \Phi(z)]^{sn_s - 1}\, dz,$$

where $\phi(z)$ and $\Phi(z)$ denote the standard normal PDF and CDF, respectively. Substituting back into Eq. 7 yields

$$\mathbb{E}[Z] = \mu + \sigma\, sn_s \int_{-\infty}^{\infty} z\ \phi(z)\,[1 - \Phi(z)]^{sn_s - 1}\, dz.$$

This is the integral form of the expectation of $\mathrm{MRP}_s$. Substituting into Eq. 6 yields

$$b_{\mathrm{MRP}_s} = \sigma\, sn_s \int_{-\infty}^{\infty} z\ \phi(z)\,[1 - \Phi(z)]^{sn_s - 1}\,dz, \tag{8}$$

which gives us the exact amount of bias of $\mathrm{MRP_s}$.

## Asymptotic Approximation of Bias

We can asymptotically approximate Eq. 8 for large values of $n_s$ using Extreme Value Theory for the minimum of $sn_s$ i.i.d. standard normal variables. Let $X_1', X_2', \dots, X_{sn_s}' \sim \mathcal{N}(0,1)$ be i.i.d., and define

$$M_{\max} = \max(X_1', \dots, X_{sn_s}'), \quad M_{\min} = \min(X_1', \dots, X_{sn_s}').$$

Since the normal distribution is symmetric,

$$M_{\min} \overset{d}{=} -M_{\max}, \quad \Rightarrow \quad \mathbb{E}[M_{\min}] = -\mathbb{E}[M_{\max}].$$

The cumulative distribution function (CDF) of the maximum is

$$P(M_{\max} \le x) = [\Phi(x)]^{sn_s},$$

and the survival function is

$$P(M_{\max} > x) = 1 - [\Phi(x)]^{sn_s}.$$

For large $n_s$, the maximum lies far in the right tail where $1 - \Phi(x)$ is small.

We will define $b$ to be the approximate location of the maximum. Specifically, we define it to be the location of the distribution where only one sample exceeds it:

$$1 - \Phi(b) = \frac{1}{sn_s}. \tag{9}$$

We will use the normal tail approximation with the Mills ratio. For large $x$,

$$1 - \Phi(x) \approx \frac{\phi(x)}{x} = \frac{1}{x\sqrt{2\pi}} e^{-x^2/2}. \tag{10}$$

Substituting Eq. 10 into Eq. 9 yields

$$\frac{1}{b\sqrt{2\pi}}e^{-b^2/2} \approx \frac{1}{sn_s}.$$

Taking natural logarithms of both sides and multiplying by -1,

$$\frac{b^2}{2} + \ln(b) + \frac{1}{2}\ln(2\pi) \approx \ln sn_s.$$

We will now find the leading-order term. For large $n_s$, $b$ is large, so $\ln(b)$ and $\ln(2\pi)$ are negligible compared with $b^2/2$. Thus, the leading order is:

$$\frac{b^2}{2} \approx \ln sn_s \quad \Rightarrow \quad b \approx \sqrt{2\ln sn_s}.$$

To obtain a more accurate approximation, we define

$$b^2 = 2\ln sn_s - c,$$

where $c$ is a small correction. Substitute into the previous logarithmic equation:

$$\frac{1}{2}(2\ln sn_s - c) + \ln(b) + \frac{1}{2}\ln(2\pi) = \ln sn_s.$$

Simplifying gives

$$-\frac{c}{2} + \ln(b) + \frac{1}{2}\ln(2\pi) = 0.$$

Using the leading-order approximation $b \approx \sqrt{2\ln sn_s}$, we have

$$-\frac{c}{2} + \frac{1}{2}\ln(2\ln sn_s) + \frac{1}{2}\ln(2\pi) \approx 0,$$

so

$$c \approx \ln(4\pi \ln sn_s).$$

Therefore,

$$b^2 \approx 2\ln sn_s - \ln(\ln n_s) - \ln(4\pi).$$

Using the first-order Taylor approximation $\sqrt{A-\varepsilon} \approx \sqrt{A} - \frac{\varepsilon}{2\sqrt{A}}$, with $A = 2\ln sn_s$ and $\varepsilon = \ln(\ln 2n) + \ln(4\pi)$, we get

$$b \approx \sqrt{2\ln sn_s} - \frac{\ln(\ln sn_s) + \ln(4\pi)}{2\sqrt{2\ln sn_s}.}$$

Simplifying yields

$$b \approx -\frac{\ln(\ln sn_s) + \ln(4\pi) - 4\ln sn_s}{2\sqrt{2\ln sn_s}.} \tag{11}$$

The limiting distribution of the normalized maximum is Gumbel with standard mean $\gamma \approx 0.5772$ (Euler–Mascheroni constant). The mean of the general Gumbel consists of location parameter $b$, and scale parameter $a$. Thus,

$$\mathbb{E}[M_{\max}] \approx b + a\gamma, \tag{12}$$

The scaling constant $a$ is asymptotically

$$a = \frac{1}{sn_s\phi(b)}.$$

Using the maximum approximation $1 - \Phi(x) \approx 1/sn_s$ and the normal tail approximation $1 - \Phi(x) \approx \phi(x)/x$, we find

$$a \approx \frac{1}{b}. \tag{13}$$

Substituting into Eq. 12 yields

$$\mathbb{E}[M_{\max}] \approx b + \frac{\gamma}{b}$$

For large $n$, $b$ grows faster while $\frac{\gamma}{b}$ shrinks, so we can approximate with

$$\mathbb{E}[M_{\max}] \approx b$$

Since $\mathbb{E}[Z'] = \mathbb{E}[M_{\min}] = -\mathbb{E}[M_{\max}]$,

$$\mathbb{E}[Z'] \approx -\sqrt{2\ln sn_s} + \frac{\ln(\ln sn_s) + \ln(4\pi)}{2\sqrt{2\ln sn_s}}.$$

For a general normal distribution $\mathcal{N}(\mu, \sigma^2)$, substituting into Eq. 7 yields,

$$\mathbb{E}[Z] \approx \mu - \sigma\sqrt{2\ln(sn_s)} + \sigma\,\frac{\ln(\ln(sn_s)) + \ln(4\pi)}{2\sqrt{2\ln(sn_s)}}.$$

Substituting into Eq. 6 yields

$$b_{\mathrm{MRP}_s} \approx \sigma \frac{\ln\left(\ln(sn_s)\right) + \ln(4\pi) - 4\ln(sn_s)}{2\sqrt{2\ln(sn_s)}}, \tag{14}$$

which gives us an asymptotic approximation of the bias of $\mathrm{MRP_s}$.

## Inconsistency of $\mathrm{MRP}_s$

We will demonstrate that $\mathrm{MRP}_s$ is not a consistent estimator. We aim to find the consistency of $Z$ when $s$ is fixed and $n_s \to \infty$. Let

$$X_{i,j} \sim \mathcal{N}(\mu, \sigma^2), \quad i = 1, \dots, n_s, \quad j = 1, \dots, s,$$

and define

$$Y_i = \min_{1 \le j \le s} X_{i,j}, \quad Z = \min_{1 \le i \le n_s} Y_i.$$

We define the standardized variables

$$X'_{i,j} = \frac{X_{i,j} - \mu}{\sigma} \sim \mathcal{N}(0,1), \quad Y'_i = \min_{1 \le j \le s} X'_{i,j}, \quad Z' = \min_{1 \le i \le n_s} Y'_i.$$

We have

$$Z = \mu + \sigma Z' \Rightarrow \mathbb{E}[Z] = \mu + \sigma \mathbb{E}[Z'].$$

The CDF of each $Y'_i$ is

$$F_{Y'}(x) = 1 - [1 - \Phi(x)]^s,$$

so the CDF of $Z'$ is

$$F_{Z'}(x) = 1 - [1 - F_{Y'}(x)]^{n_s} = 1 - ((1 - \Phi(x))^s)^{n_s} = 1 - (1 - \Phi(x))^{sn_s}.$$

We will now look at the limit as $n_s \to \infty$ with $s$ fixed. For any finite $x$,

$$\Pr(Z' > x) = (1 - \Phi(x))^{sn_s} = ((1 - \Phi(x))^s)^{n_s}.$$

Since $0 < (1 - \Phi(x))^s < 1$ for finite $x$, we have

$$\lim_{n_s \to \infty} \Pr(Z' > x) = 0 \quad \Rightarrow \quad \Pr(Z' \le x) \to 1.$$

Therefore,

$$Z' \to -\infty$$

We have

$$Z = \mu + \sigma Z' \to -\infty$$

Because $Z = \text{MRP}_s$, we have

$$\lim_{n_s \to \infty} \text{MRP}_s(r) \to -\infty.$$

So $Z$ does not converge in probability to $\mathbb{E}[X]$, or any finite number. Therefore, $Z$ is not a consistent estimator of $\mathbb{E}[Z]$.

Even though $Z$ diverges, Extreme Value Theory gives a nondegenerate limit after centering and scaling. We have

$$\frac{Z' + b}{a} \to -G$$

where $G$ is a Gumbel random variable with mean $\gamma$, and $a$ is defined in Eq. 11 and $b$ is defined in Eq. 13. Thus,

$$\frac{Z - (\mu - \sigma b)}{\sigma a}$$

converges in distribution to $-G$. This method also shows that $Z$ is not a consistent estimator of $\mathbb{E}[X]$.

# Appendix B: Empirical Implementation and Sensitivity Tests

We describe the empirical implementation and provide complete sensitivity tests for all factors.

## Empirical Implementation

The factor returns data was collected from https://jkpfactors.com/?dataFrequency=daily. The data was truncated starting at 1980. $\text{MRP}_1$ was calculated using Eq. 1 and Eq. 2. The sensitivity analysis was performed by using different lookbacks and $d$ values to calculate Eq. 2.

## Complete Sensitivity Analysis

While Exhibit 5 and Exhibit 6 show aggregated sensitivity analysis results, we also calculate full sensitivity tests for each factor.

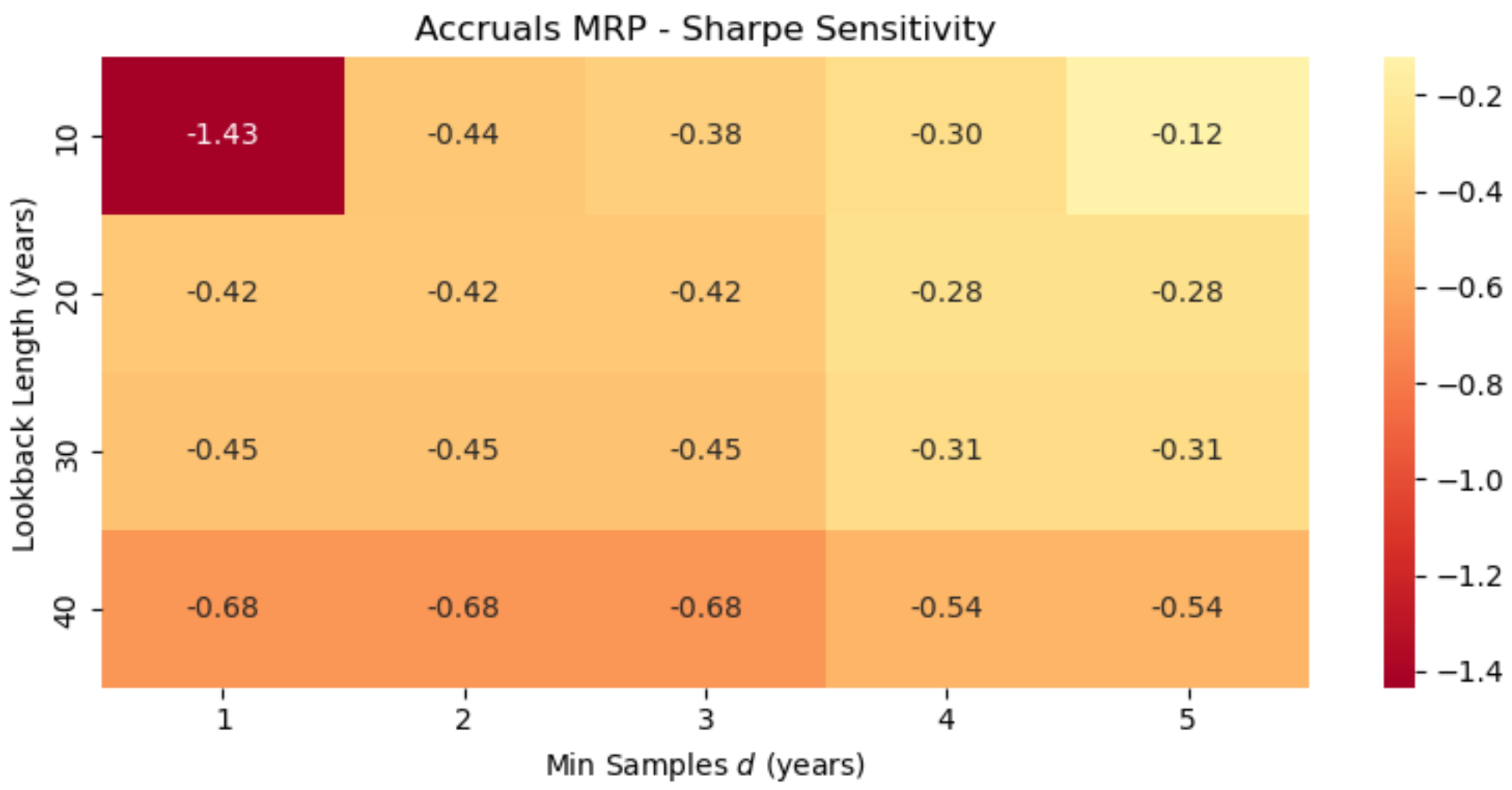

Exhibit 8: MRP – Sharpe Sensitivity of Accruals

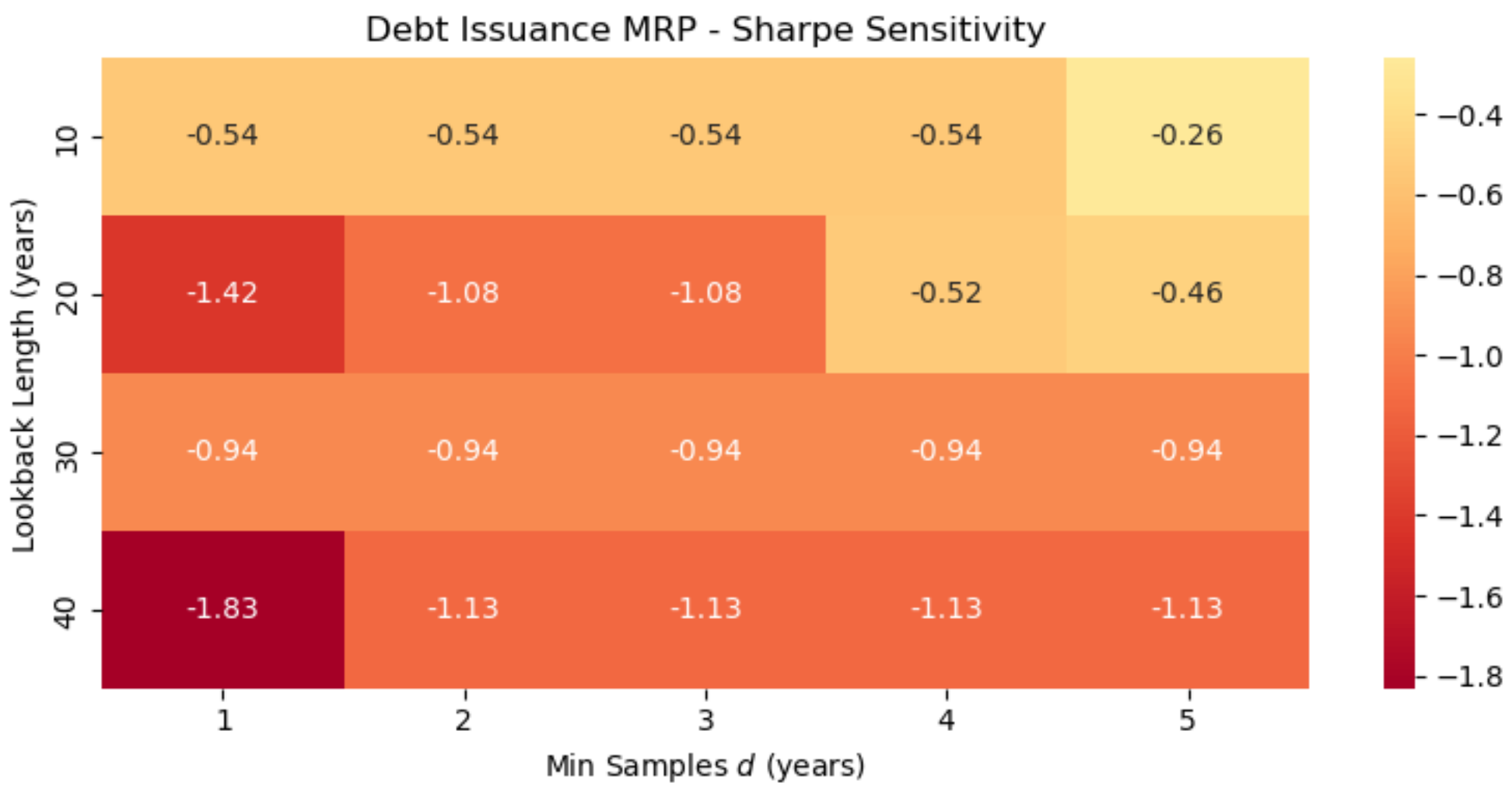

Exhibit 9: MRP – Sharpe Sensitivity of Debt Issuance

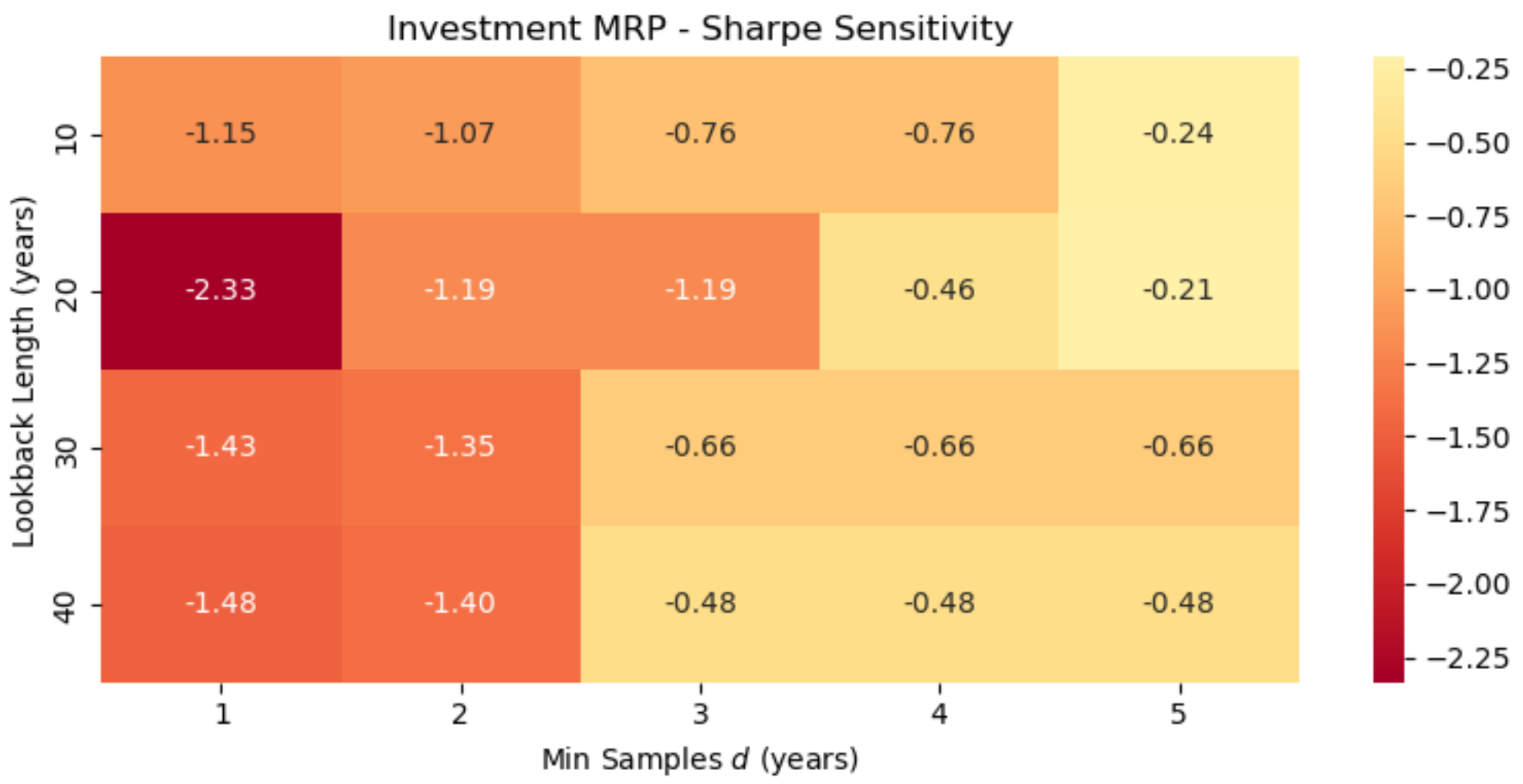

Exhibit 10: MRP – Sharpe Sensitivity of Investment

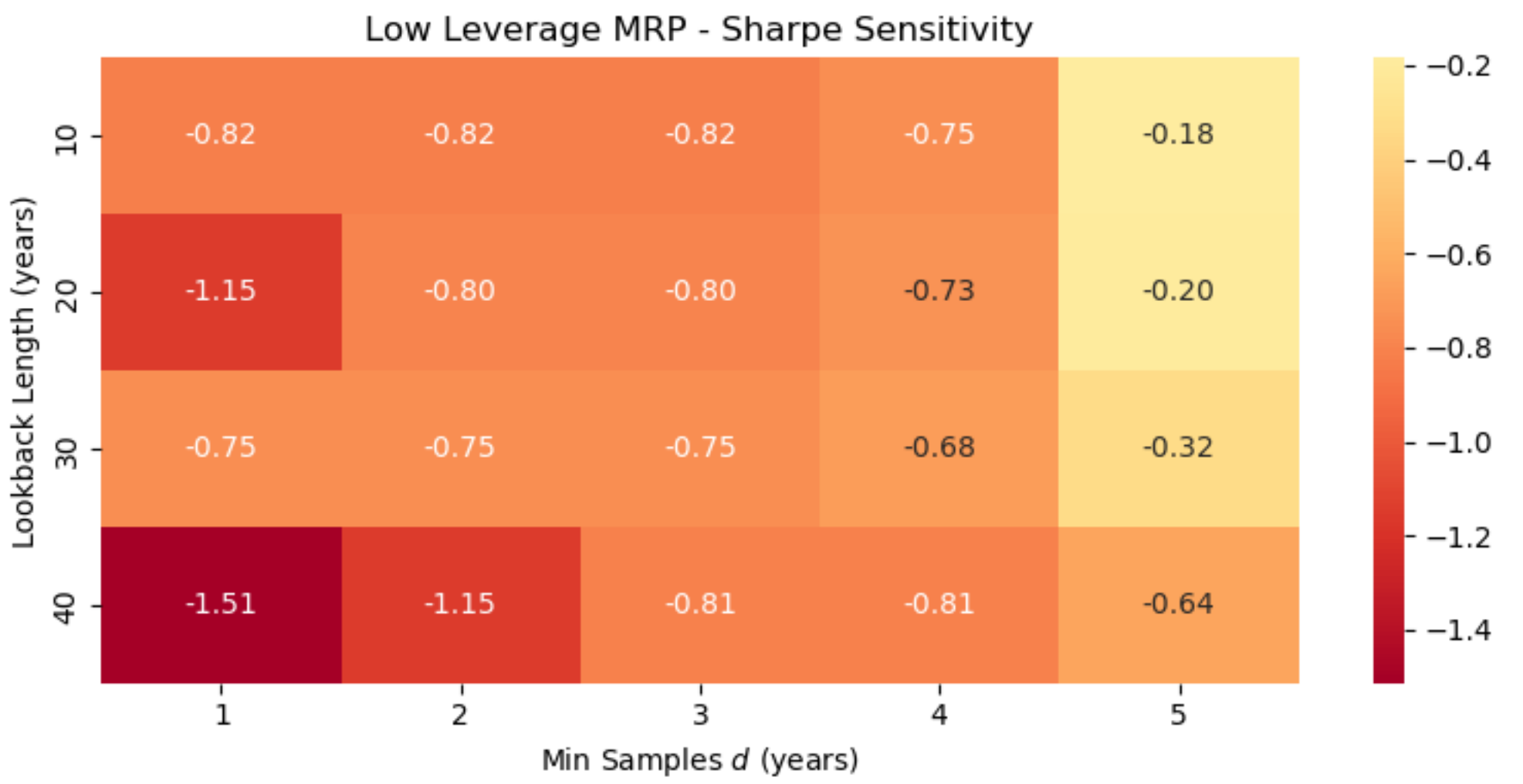

Exhibit 11: MRP – Sharpe Sensitivity of Low Leverage

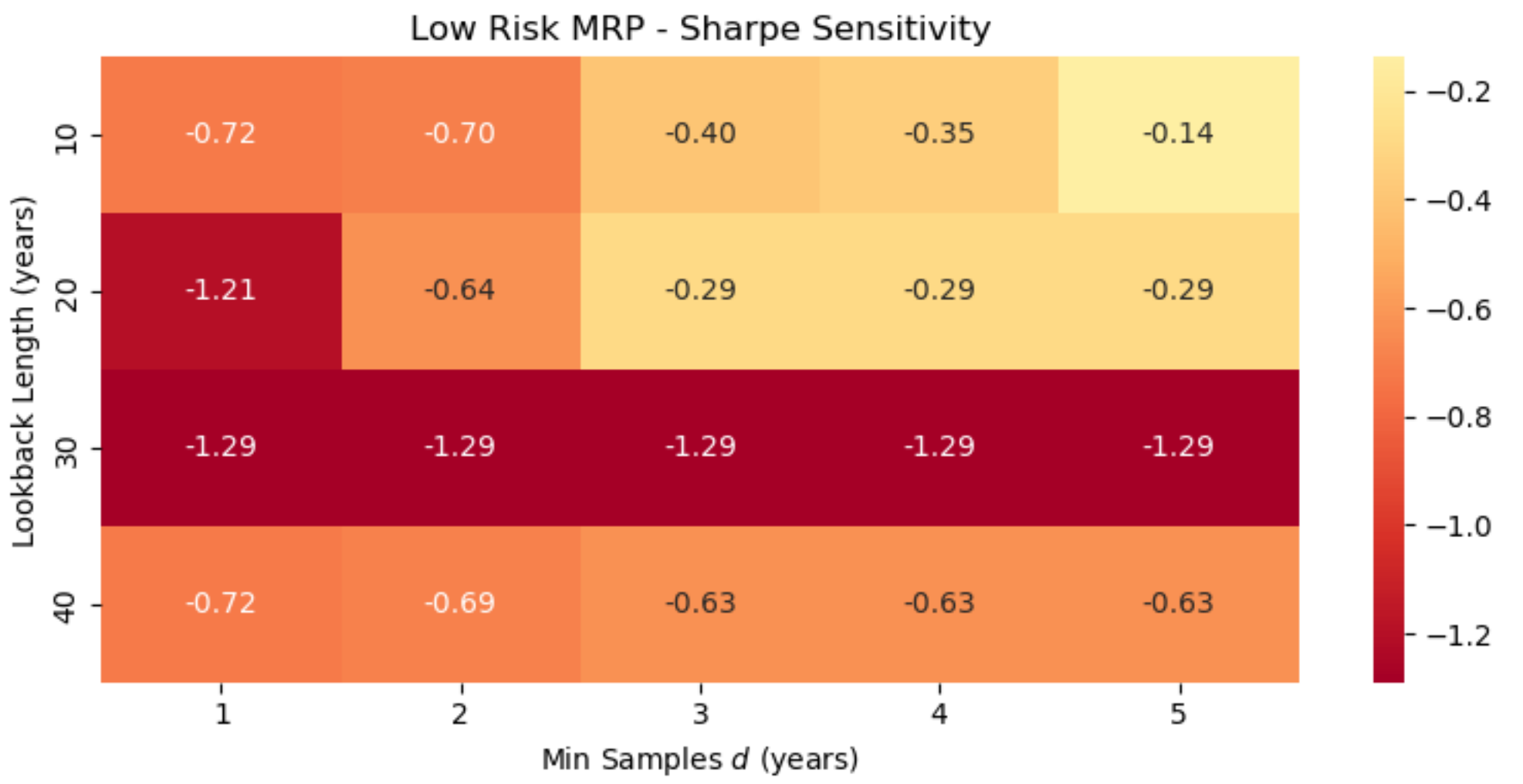

Exhibit 12: MRP – Sharpe Sensitivity of Low Risk

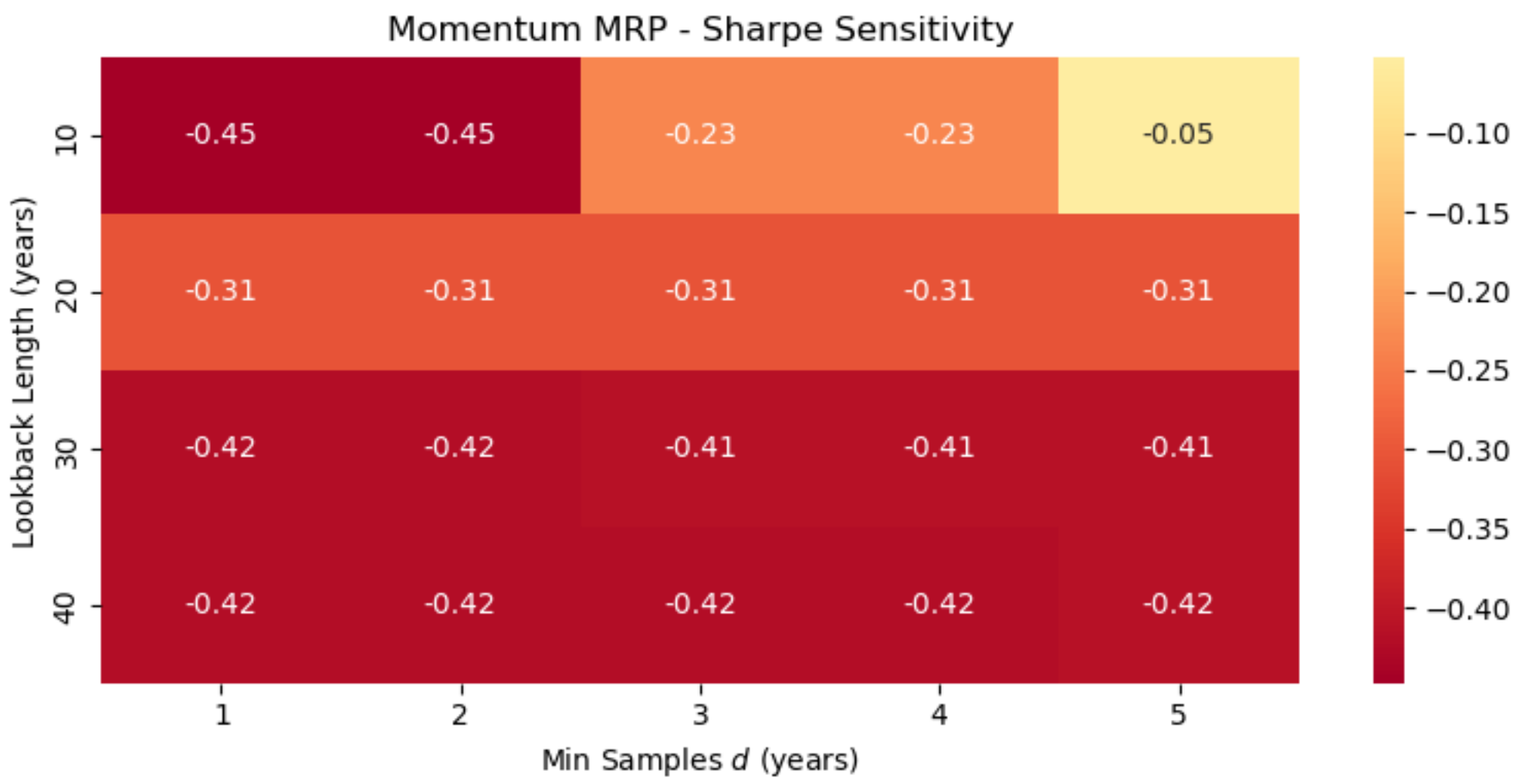

Exhibit 13: MRP – Sharpe Sensitivity of Momentum

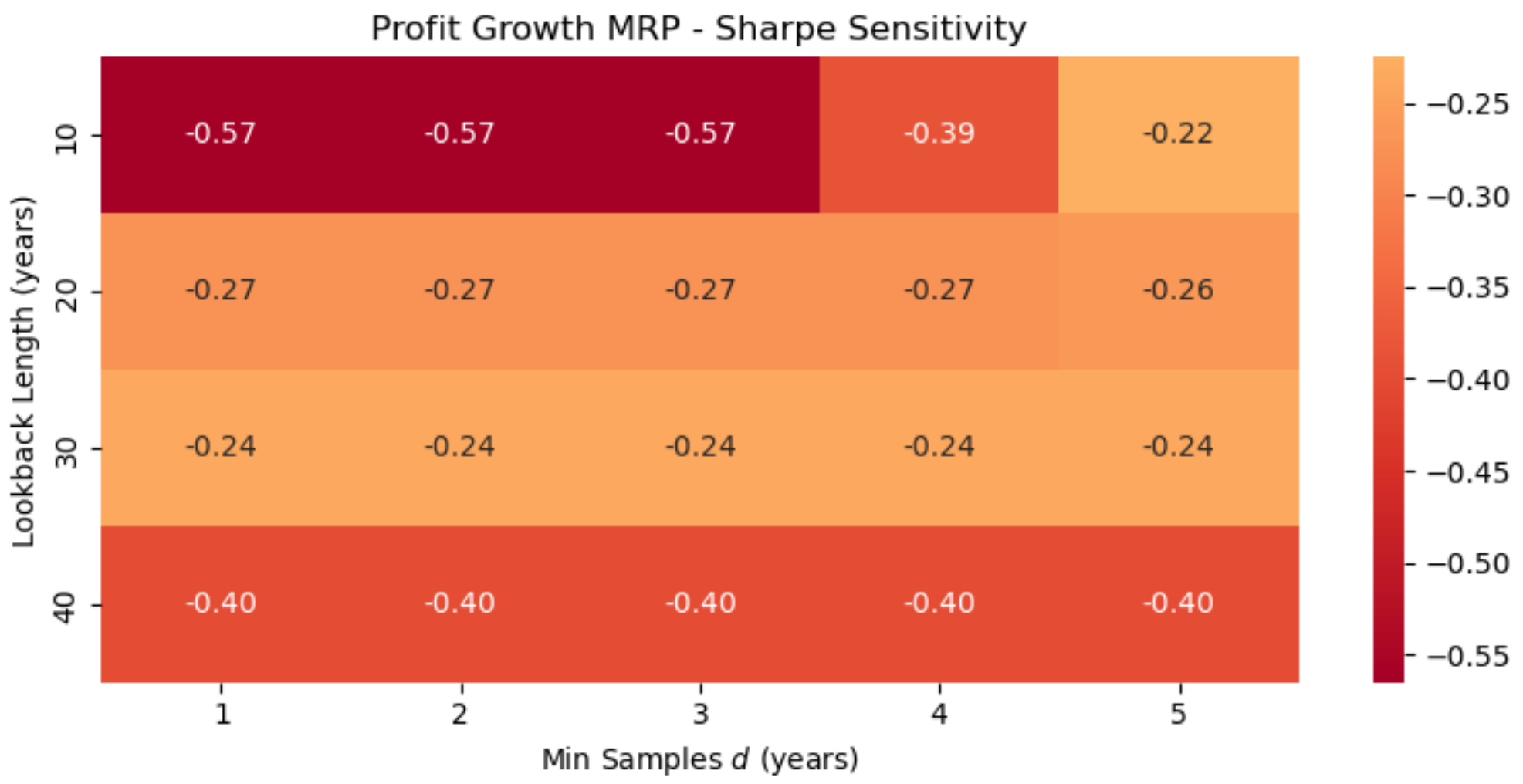

Exhibit 14: MRP – Sharpe Sensitivity of Profit Growth

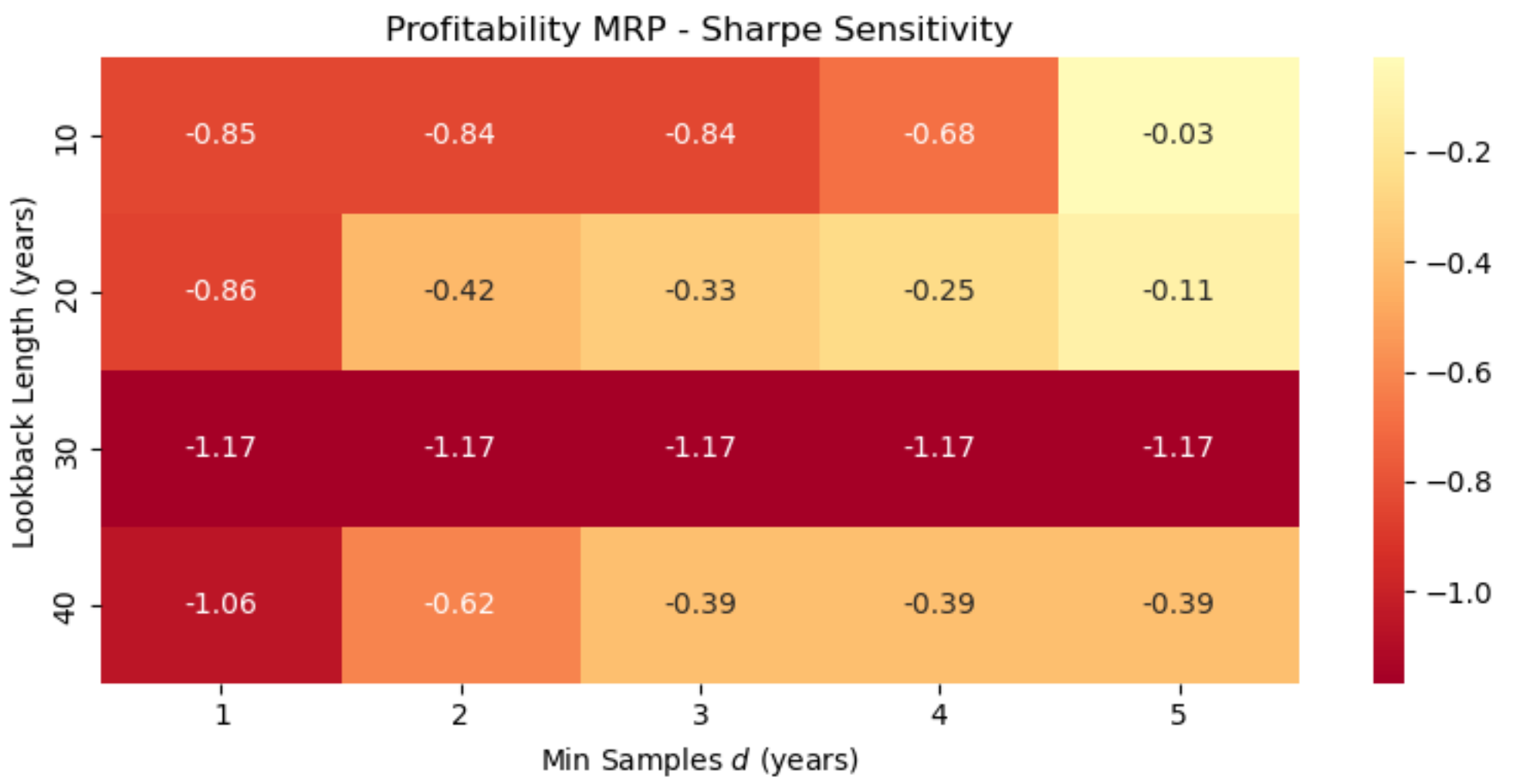

Exhibit 15: MRP – Sharpe Sensitivity of Profitability

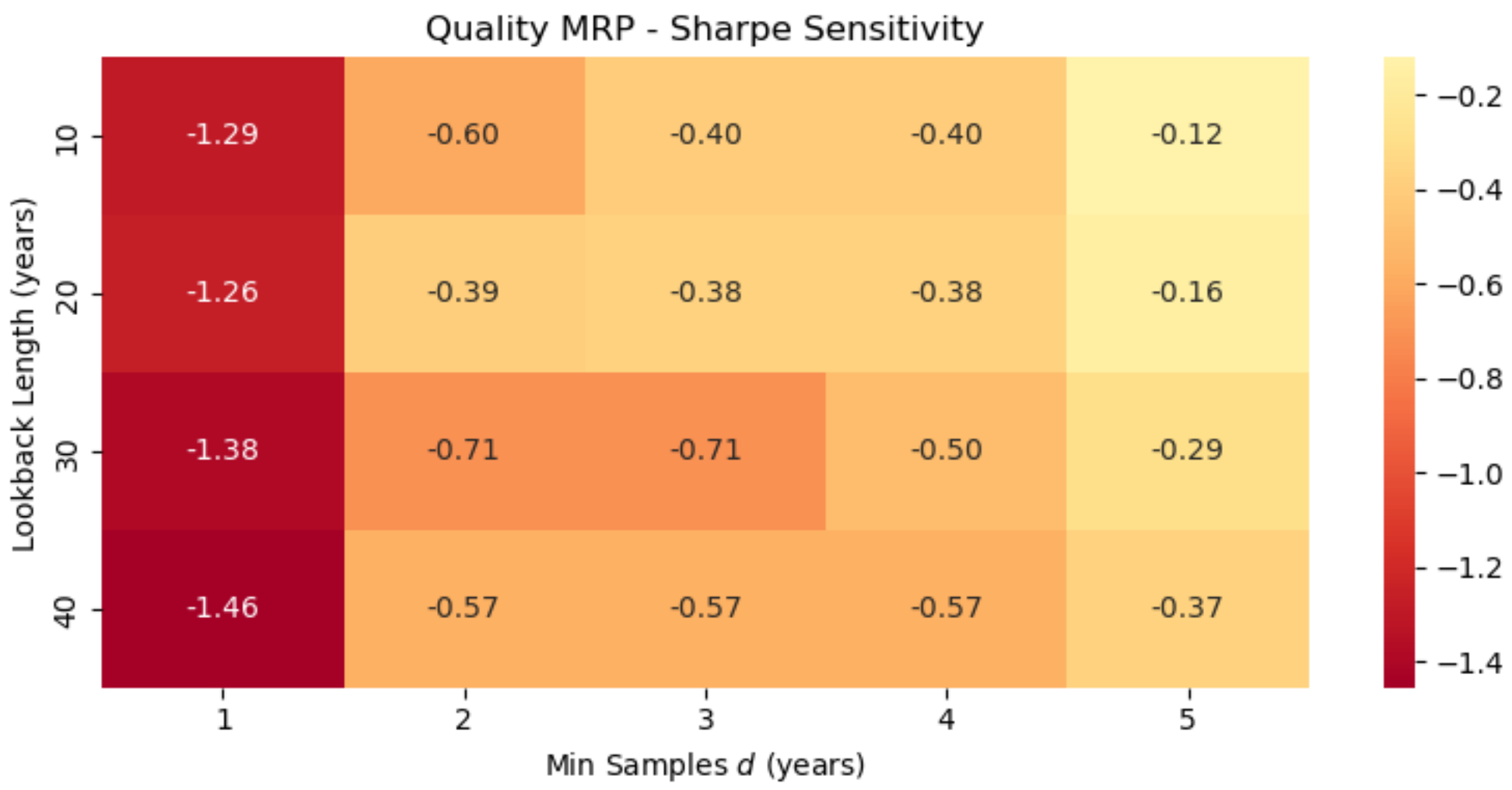

Exhibit 16: MRP – Sharpe Sensitivity of Quality

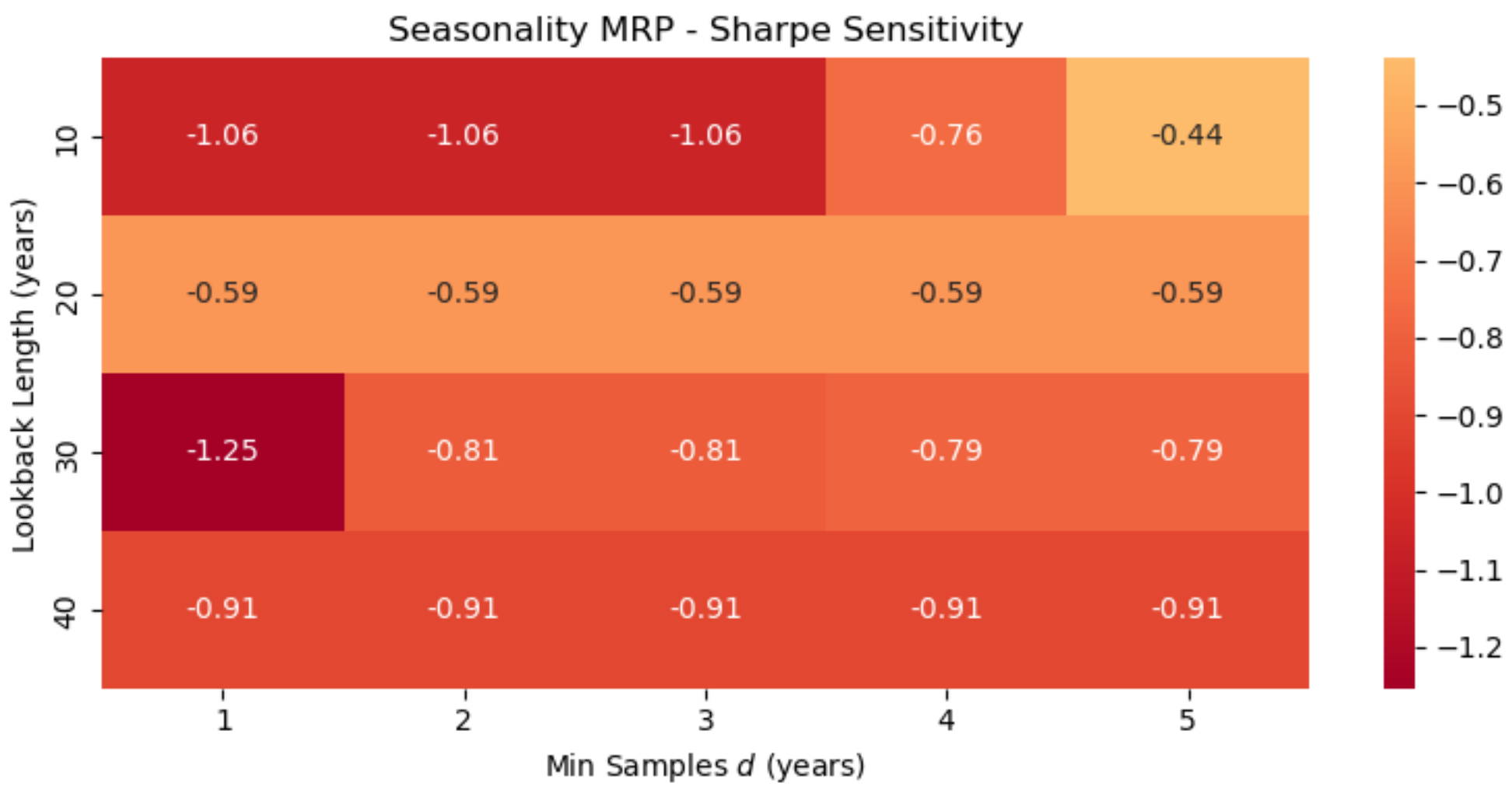

Exhibit 17: MRP – Sharpe Sensitivity of Seasonality

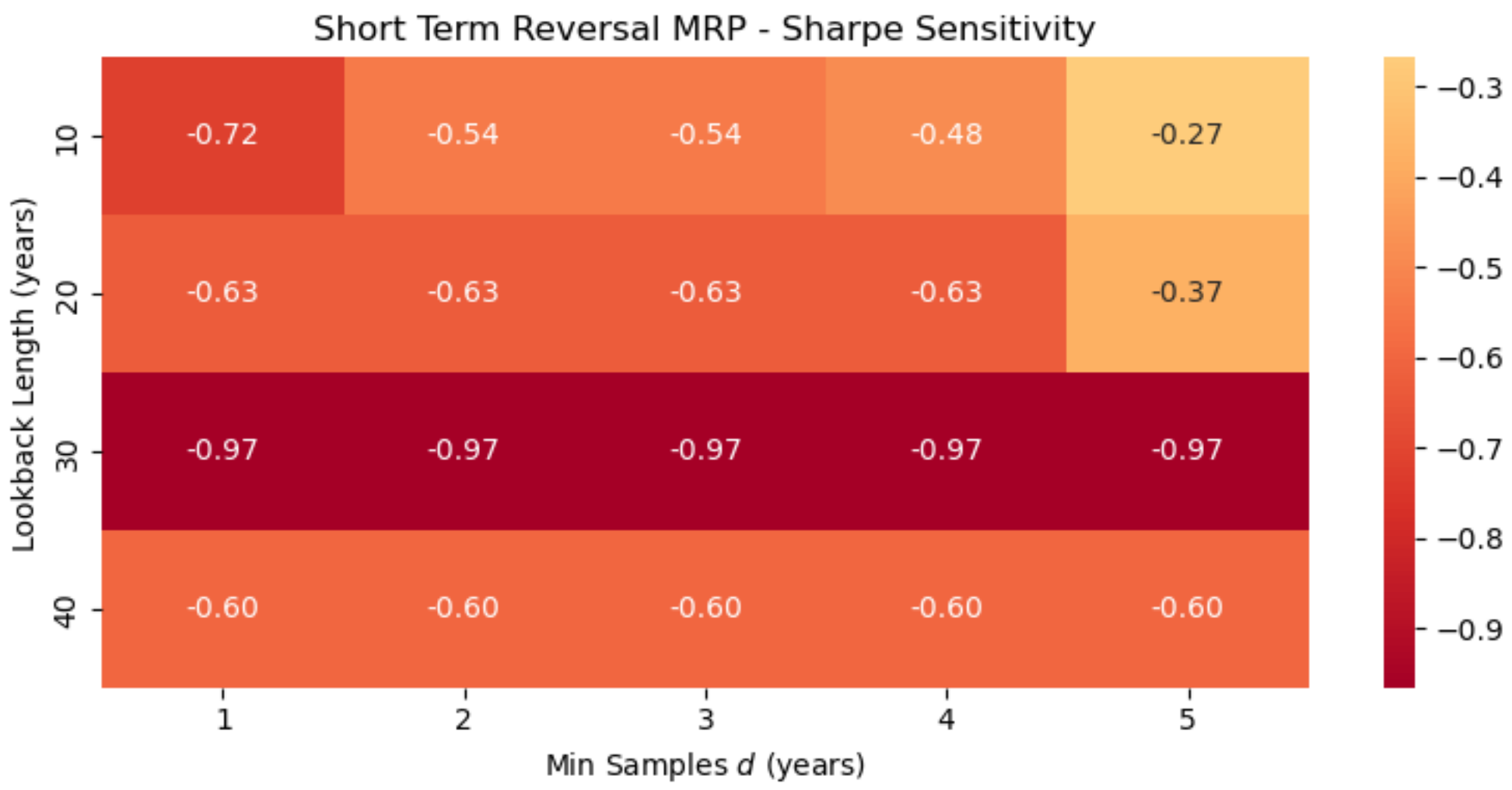

Exhibit 18: MRP – Sharpe Sensitivity of Short-term Reversal

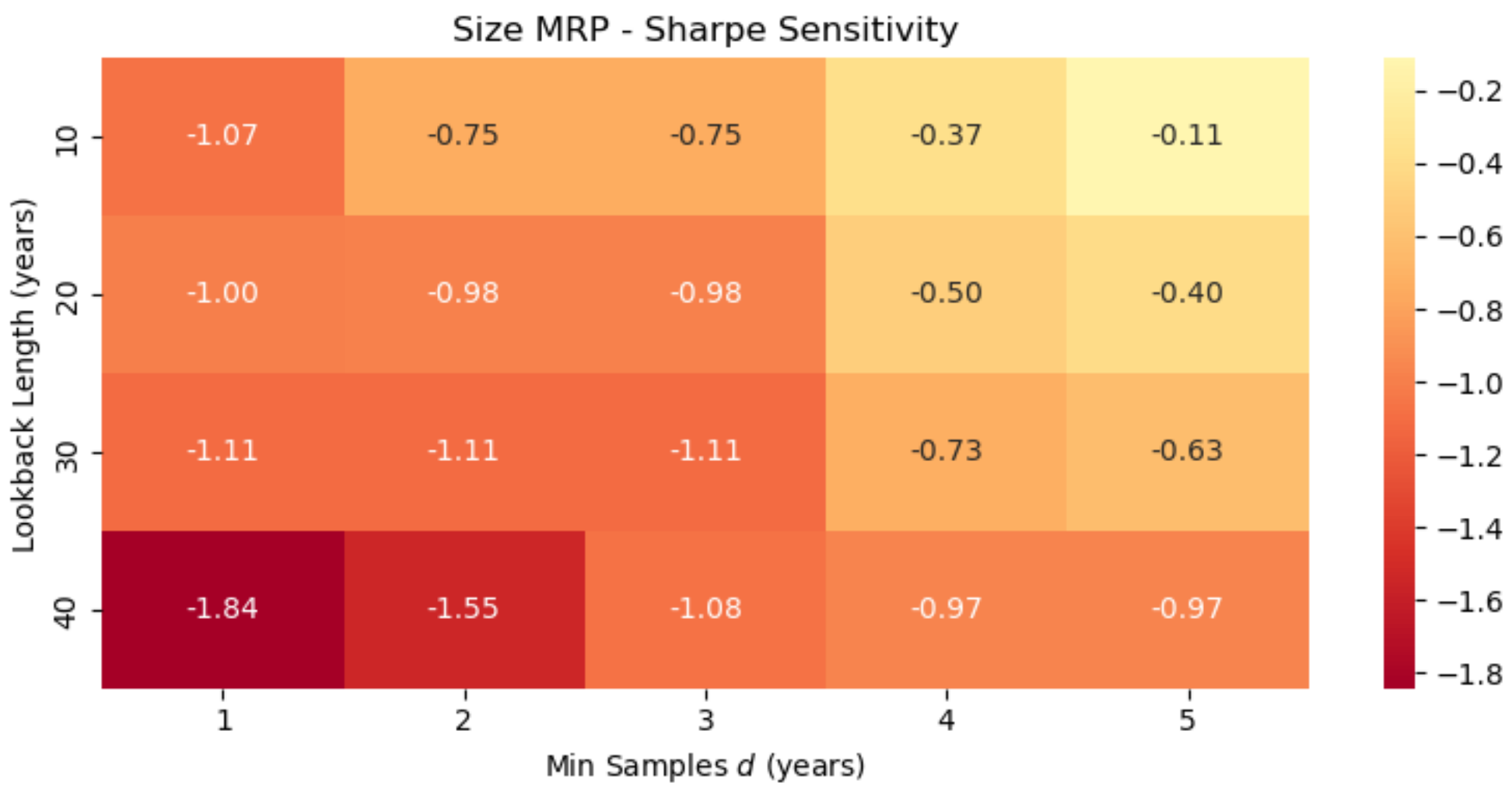

Exhibit 19: MRP – Sharpe Sensitivity of Size

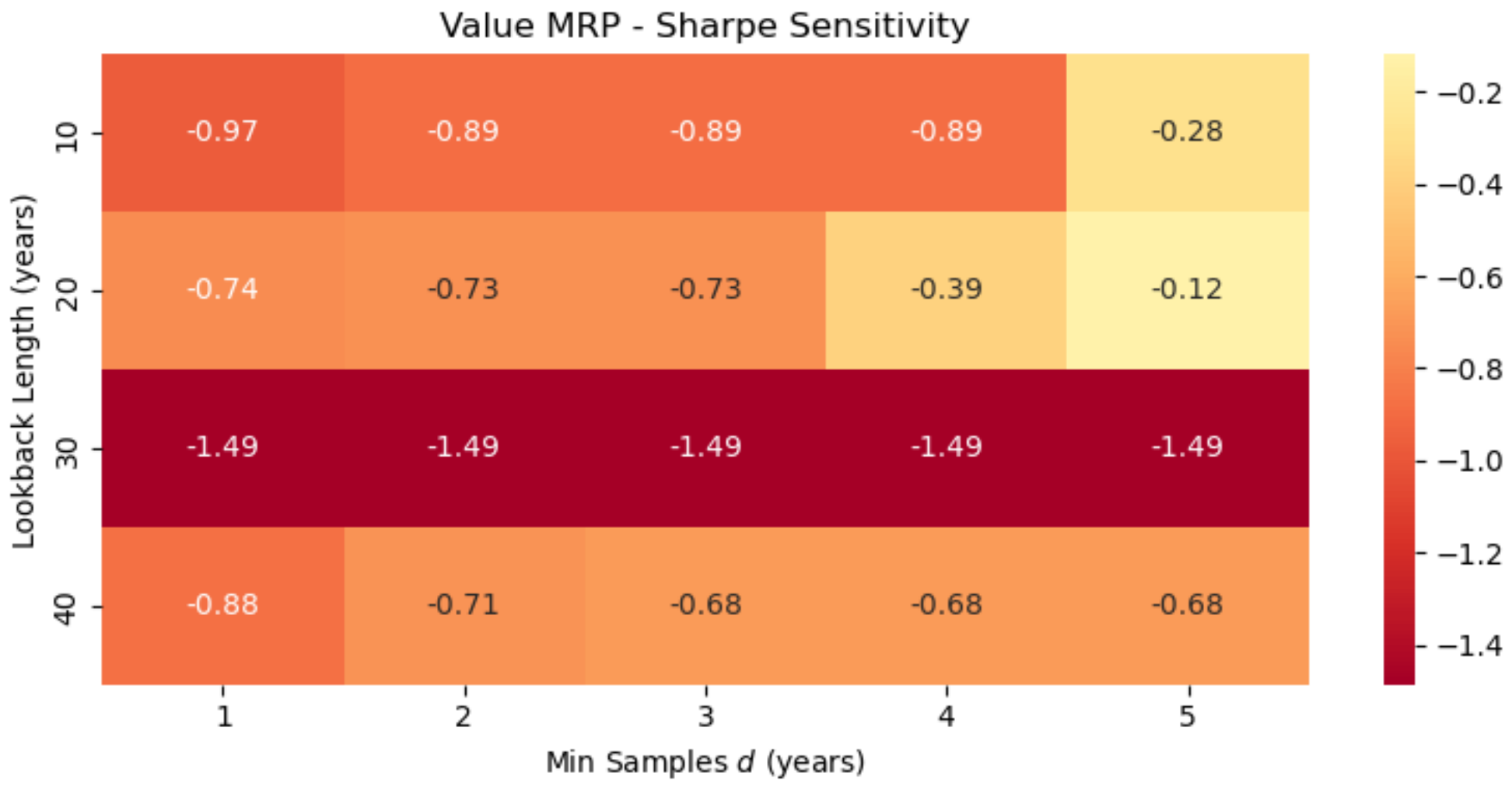

Exhibit 20: MRP – Sharpe Sensitivity of Value